\documentclass[aps, prd, twocolumn, nofootinbib, showpacs]{revtex4}

\usepackage{amsfonts,amsmath,amsthm,amssymb,graphicx}

\begin{document}

\title{Quantum modes in DBI inflation: exact solutions and constraints from vacuum selection}

\author{William H.  Kinney} 
\email{whkinney@buffalo.edu}
\affiliation{Perimeter Institute for Theoretical Physics, 31 Caroline Street North, Waterloo, Ontario, Canada N2L 2Y5}
\affiliation{Dept. of Physics, University at Buffalo, the  State University of New York, Buffalo, NY 14260-1500}
\author{Konstantinos Tzirakis} 
\email{ct38@buffalo.edu}
\affiliation{Dept. of Physics, University at Buffalo, the  State University of New York, Buffalo, NY 14260-1500}

\begin{abstract}
We study a two-parameter family of exactly solvable inflation models with variable sound speed, and derive a corresponding exact expression for the spectrum of curvature perturbations. We generalize this expression to the slow roll case, and derive an approximate expression for the scalar spectral index valid to second order in slow roll. We apply the result to the case of DBI inflation, and show that for certain choices of slow roll parameters, the Bunch-Davies limit (a) does not exist, or (b) is sensitive to stringy physics in the bulk, which in principle can have observable signatures in the primordial power spectrum. 
\end{abstract}


\maketitle

\section{Introduction}
\label{sec:introduction}

Inflationary cosmology \cite{Guth:1980zm,Linde:1981mu,Albrecht:1982wi} is an extremely successful phenomenological paradigm, making quantitative predictions which have been strongly supported by recent data \cite{Spergel:2006hy,Alabidi:2006qa,Seljak:2006bg,Kinney:2006qm,Martin:2006rs}. However, no compelling fundamental explanation for inflation has yet been proposed. An obvious place to search for a fundamental theory of inflation is within the ``landscape'' of string theory, which predicts a plethora of scalar fields associated with the compactification of extra dimensions and the configuration of lower-dimensional ``branes'' moving in a higher-dimensional bulk space. Recent developments in string theory have produced a number of phenomenologically viable stringy inflation models such at the KKLMMT scenario \cite{Kachru:2003sx}, Racetrack Inflation \cite{BlancoPillado:2004ns}, and Roulette Inflation \cite{Bond:2006nc}. The DBI scenario \cite{Silverstein:2003hf} has attracted particular interest because of the novel feature that slow roll can be achieved through a low sound speed instead of from dynamical friction due to expansion. This behavior introduces novel phenomenology, in particular significant non-Gaussianity \cite{Alishahiha:2004eh,Chen:2006nt,Spalinski:2007qy,Bean:2007eh,LoVerde:2007ri}, which can potentially be used to distinguish DBI inflation from other scenarios. In the DBI scenario, the field $\phi$ responsible for inflation (the {\it inflaton}) is the degree of freedom associated with a $3+1$-dimensional brane with metric $ds_4^2$ moving in a six-dimensional ``throat'', with a metric of the form \cite{Klebanov:2000hb}
\begin{equation}
\label{eq:DBImetric}
ds^2_{10} = h^2\left(r\right) ds^2_4 + h^{-2}\left(r\right) \left(d r^2 + r^2 ds^2_{X_5}\right).
\end{equation}
The field $\phi$ is simply related to the coordinate in the throat $r$ as $\phi = \sqrt{T_3} r$, where the brane tension  $T_3$ depends on the string scale $m_s$ and the string coupling $g_s$ as \cite{Lidsey:2007gq}
\begin{equation}
T_3 = \frac{m_s^4}{\left(2 \pi\right)^3 g_s}.
\end{equation}
The Lagrangian for the inflaton $\phi$ can be shown to be of the form
\begin{equation}
\label{eq:DBIlagrangian}
{\cal L} = - f^{-1}\left(\phi\right) \sqrt{1 + f\left(\phi\right) g^{\mu\nu} \partial_\mu \phi \partial_\nu \phi} + f^{-1}\left(\phi\right) - V\left(\phi\right), 
\end{equation}
where $V\left(\phi\right)$ is an arbitrary potential, and the inverse brane tension $f\left(\phi\right)$ is given in terms of the warp factor $h\left(\phi\right)$ by
\begin{equation}
\label{eq:warpfactor}
f\left(\phi\right) = \frac{1}{T_3 h^4\left(\phi\right)}.
\end{equation}
We take the four-dimensional metric to be of the Friedmann-Robertson-Walker form $g_{\mu \nu} = {\rm diag}\left(-1,a^2(t),a^2(t),a^2(t)\right)$. 

Although the Lagrangian (\ref{eq:DBIlagrangian}) is fundamentally derived from stringy physics, from the point of view of cosmology it can be treated simply as a phenomenological model, and its observational consequences can be determined from the specification of the field Lagrangian. In this sense, the DBI scenario is  a special case of a larger class of inflationary models with non-canonical Lagrangians and a time-dependent speed of sound, first proposed by Armendariz-Picon, Damour, and Mukhanov under the name k-inflation \cite{ArmendarizPicon:1999rj}. Following this point of view, in Section \ref{sec:background} of this paper, we consider a set of exact solutions for the background evolution of the inflaton in a general k-inflation scenario, first derived by Chimento and Lazkoz \cite{Chimento:2007es} and by Spalinski \cite{Spalinski:2007qy,Spalinski:2007un}. These solutions are generalizations of power-law inflation in the case of a canonical inflaton field, which has an exactly solvable perturbation spectrum. In Section \ref{sec:modes}, we obtain exact solutions for the scalar perturbation spectrum. We use the exact solution to obtain a general expression for the spectral index of scalar perturbations valid to second-order in the slow roll parameters which is the analog of the expression obtained by Stewart and Lyth \cite{Stewart:1993bc} for the case of a canonical inflaton. In Section \ref{sec:vacuum}, we discuss these solutions in the context of the DBI scenario. In the case of DBI, the Lagrangian (\ref{eq:DBIlagrangian}) is only well-defined when the brane is well within the throat geometry, where the metric is of the form (\ref{eq:DBImetric}). Outside of the throat, the bulk spacetime will have a more general metric. This places strong constraints on the DBI scenario, and introduces the possibility that the vacuum state for quantum modes in inflation may be sensitive to the stringy physics in the bulk, with corresponding observational signatures in the primordial density perturbations. Section \ref{sec:Conclusions} presents a summary and conclusions.  

\section{Exact background solutions}
\label{sec:background}

In this section, we generate a two-parameter family of exact solutions for DBI inflation. 
We begin with the generalization of Peiris {\it et al.} to the inflationary flow formalism \cite{Peiris:2007gz}. In addition to the canonical flow parameters $\epsilon$, $\eta$, and ${}^\ell \lambda$, which parameterize the time dependence of the Hubble length, DBI inflation models (as well as any other models with a varying speed of sound such as k-inflation \cite{ArmendarizPicon:1999rj}), require a second hierarchy of parameters $s$, $\rho$, and ${}^\ell \alpha$ which parameterize the time dependence of the sound horizon. For a Lagrangian of the form (\ref{eq:DBIlagrangian}), the speed of sound is given by $c_S = 1 / \gamma\left(\phi\right)$, where
\begin{equation}
\gamma\left(\phi\right) = \frac{1}{\sqrt{1 - f\left(\phi\right) \dot\phi^2}}.
\end{equation}
The flow hierarchy defined by Peiris, {\it et al.} is:
\begin{eqnarray}
\label{eq:flowparams}
\epsilon\left(\phi\right) &\equiv& \frac{2 M_P^2}{\gamma\left(\phi\right)} \left(\frac{H'\left(\phi\right)}{H\left(\phi\right)}\right)^2,\cr
\eta\left(\phi\right) &\equiv& \frac{2 M_P^2}{\gamma\left(\phi\right)} \frac{H''\left(\phi\right)}{H\left(\phi\right)},\cr
s\left(\phi\right) &\equiv& \frac{2 M_P^2}{\gamma\left(\phi\right)} \frac{H'\left(\phi\right)}{H\left(\phi\right)} \frac{\gamma'\left(\phi\right)}{\gamma\left(\phi\right)},\cr
\rho\left(\phi\right) &\equiv& \frac{2 M_P^2}{\gamma\left(\phi\right)}  \frac{\gamma''\left(\phi\right)}{\gamma\left(\phi\right)},\cr
{}^\ell \lambda\left(\phi\right) &\equiv& \left(\frac{2 M_P^2}{\gamma\left(\phi\right)}\right)^{\ell} \left(\frac{H'\left(\phi\right)}{H\left(\phi\right)}\right)^{\ell - 1} \frac{1}{H\left(\phi\right)} \frac{d^{\ell + 1} H\left(\phi\right)}{d \phi^{\ell + 1}},\cr
{}^\ell \alpha\left(\phi\right) &\equiv& \left(\frac{2 M_P^2}{\gamma\left(\phi\right)}\right)^{\ell} \left(\frac{H'\left(\phi\right)}{H\left(\phi\right)}\right)^{\ell - 1} \frac{1}{\gamma\left(\phi\right)} \frac{d^{\ell + 1} \gamma\left(\phi\right)}{d \phi^{\ell + 1}},
\end{eqnarray}
where $\ell = 2,\ldots,\infty$ is an integer index.\footnote{Our notation differs slightly from that of Peiris {\it et al.}, who use $\kappa$ in place of s.} This reduces to the usual canonical flow hierarchy \cite{Kinney:2002qn} for $\gamma = 1$. We define $N$ to be the number of e-folds before the end of inflation
\begin{equation}
\label{eq:numefolds}
N \equiv - \int{H}{dt} = \frac{1}{\sqrt{2 M_P^2}} \int_{\phi_e}^{\phi}{\sqrt{\frac{\gamma\left(\phi\right)}{\epsilon\left(\phi\right)}} d\phi},
\end{equation}
where $\phi = \phi_e$ is the end of inflation, so that $N = 0$ at the end of inflation, and increases as we go backward in time. The flow parameters (\ref{eq:flowparams}) are related by a series of first-order {\it flow equations},
\begin{eqnarray}
\label{eq:flowequations}
\epsilon &=& \frac{1}{H}\frac{d H}{d N},\cr
\frac{d \epsilon}{d N} &=& \epsilon\left(2 \eta - 2 \epsilon - s\right),\cr
\frac{d \eta}{d N} &=& -\eta\left(\epsilon + s\right) + {}^2 \lambda,\cr
\frac{d {}^\ell \lambda}{d N} &=& -{}^\ell \lambda \left[\ell \left(s + \epsilon\right) - \left(\ell - 1\right) \eta\right] + {}^{\ell + 1}\lambda,\cr
s &=& \frac{1}{\gamma}\frac{d \gamma}{d N},\cr
\frac{d s}{d N} &=& -s \left(2 s + \epsilon - \eta\right) + \epsilon \rho,\cr
\frac{d \rho}{d N} &=& -2 \rho s + {}^2 \alpha,\cr
\frac{d {}^\ell \alpha}{d N} &=& -{}^\ell \alpha \left[\left(\ell + 1\right) s + \left(\ell - 1\right) (\epsilon-\eta)\right] + {}^{\ell + 1} \alpha.
\end{eqnarray}
Taken to infinite order, this set of differential equations completely specifies the evolution of the spacetime, via the functions $H\left(\phi\right)$ and $\gamma\left(\phi\right)$. These functions can be related to the inflationary dynamics using the Hamilton-Jacobi equations \cite{Spalinski:2007kt}
\begin{eqnarray}
\label{eq:Hamjacobi 1}
\dot \phi=-\frac{2M_P^2}{\gamma(\phi)}H'(\phi)&&\cr
3M_P^2H^2(\phi)-V(\phi)&=&\frac{\gamma(\phi)-1}{f(\phi)},
\end{eqnarray}
where
\begin{equation}
\label{eq:Hamjacobi 2}
\gamma=\sqrt{1+4M_P^2f(\phi)H'(\phi)}.
\end{equation}
Given a solution to the flow equations, it is always possible to unambiguously derive the forms of the inverse brane tension $f\left(\phi\right)$ and the potential $V\left(\phi\right)$ as follows:
\begin{eqnarray}
\label{eq:Hamjacobi}
f\left(\phi\right) &=& \frac{1}{2 M_{P}^2 H^2\epsilon}\left(\frac{\gamma^2 - 1}{\gamma}\right)\cr
V\left(\phi\right) &=& 3 M_P^2 H^2 \left(1 - \frac{2\epsilon}{3} \frac{\gamma}{\gamma+1}\right).
\end{eqnarray}

One solution of particular interest is the case for which all the flow parameters are constant:
\begin{equation}
\frac{d \epsilon}{d N} = \frac{d \eta}{d N} = \frac{d {}^\ell \lambda}{d N} = 0,
\end{equation}
and
\begin{equation}
\frac{d s}{d N} = \frac{d \rho}{d N} = \frac{d {}^\ell \alpha}{d N} = 0.
\end{equation}
From $\epsilon = {\rm const.}$, we immediately have that the Hubble parameter is exponential in $N$,
\begin{equation}
H \propto e^{\epsilon N},
\end{equation}
and from $s = {\rm const.}$, we have that the speed of sound $c_S = \gamma^{-1}$ is also exponential in $N$,
\begin{equation}
\label{eq:cS}
c_S \propto e^{-s N}.
\end{equation}
Once the values of $\epsilon$ and $s$ are fixed, we can obtain full solutions to the system of flow equations (\ref{eq:flowequations}) iteratively, obtaining {\it exact} solutions for the field and metric evolution during inflation. 

\subsection{Case 1: $\epsilon = s = 0$}
The simplest class of exact solutions is when $\epsilon$ vanishes, which corresponds to constant Hubble parameter $H =  {\rm const.}$, so that the expansion is exponential,
\begin{equation}
a\left(t\right) \propto e^{H t}.
\end{equation}
From Eqs. (\ref{eq:flowparams}), this requires  that $s$ vanishes, which corresponds to constant sound speed $c_S = {\rm const.}$ In this case, the sound horizon $c_S H^{-1}$ has constant length, and this is just the trivial solution of a decoupled scalar field with speed of sound $c_S$ evolving in a de Sitter background.  We will not consider this case further here.

\subsection{Case 2: $\epsilon \neq 0$, $s = 0$}
In this case, the speed of sound $c_S$ remains constant, but the Hubble parameter varies in time. The flow equations (\ref{eq:flowequations}) are solved by
\begin{eqnarray}
\eta &=& \epsilon,\cr
{}^\ell \lambda &=& \epsilon^\ell,
\end{eqnarray}
which we can recognize as the power-law fixed point in the flow space \cite{Kinney:2002qn}. From 
\begin{equation}
\epsilon \equiv \frac{2 M_P^2}{\gamma}\left(\frac{H'}{H}\right)^2 = {\rm const.},
\end{equation}
we have the following solution for $\dot\phi > 0$,
\begin{equation}
H\left(\phi\right) = H_0 \exp{\left(- \sqrt{\frac{\gamma \epsilon}{2 M_P^2}} \phi\right)}.
\end{equation}
Here we use the sign convention that $\sqrt{\epsilon}$ is always positive. 
From Eqs. (\ref{eq:Hamjacobi}), the potential $V\left(\phi\right)$ is then
\begin{equation}
\label{eq:Vscaling}
V\left(\phi\right) = 3 M_P^2 H_0^2 \left(1 - \frac{2\epsilon}{3}\frac{\gamma}{\left(1 + \gamma\right)}\right) \exp{\left(- \sqrt{\frac{2 \gamma \epsilon}{M_P^2}} \phi \right)},
\end{equation}
and $f(\phi)$ is given by
\begin{equation}
\label{eq:fscaling}
f\left(\phi\right) = \frac{1}{2 M_P^2 H_0^2 \epsilon} \left(\frac{\gamma^2 - 1}{\gamma}\right) \exp{\left(\sqrt{\frac{2 \gamma \epsilon}{M_p^2}} \phi\right)}.
\end{equation}
The case $\gamma = 1$ is the familiar canonical power-law inflation case. For $\gamma \neq 1$, the scale factor also evolves as a power law, as can be seen from $H \equiv \left(\dot a / a\right)$, and Eq. (\ref{eq:Hamjacobi 1}) expressed in terms of $\epsilon$ as follows:
\begin{equation}
\dot\phi = \sqrt{\frac{2 M_P^2 \epsilon}{\gamma}} H = H_0 \sqrt{\frac{2 M_P^2 \epsilon}{\gamma}} \exp{\left(- \sqrt{\frac{\gamma \epsilon}{2 M_P^2}} \phi\right)}.
\end{equation}
We then have
\begin{equation}
dt = H_0^{-1} \sqrt{\frac{\gamma}{2 M_P^2 \epsilon}} \exp{\left(\sqrt{\frac{\gamma \epsilon}{2 M_P^2}} \phi\right)} d \phi,
\end{equation}
with solution
\begin{equation}
H t = \frac{1}{\epsilon},
\end{equation}
and
\begin{equation}
a\left(t\right) \propto t^{1 / \epsilon}.
\end{equation}
This solution was also obtained by Spalinski \cite{Spalinski:2007un}, and represents a rescaling of the canonical power-law inflation case. 

\subsection{Case 3: $\epsilon \neq 0$, $s \neq 0$}

For $\epsilon$ and $s$ both nonzero, it is straightforward to show that the following two-parameter family of solutions satisfies the flow equations (\ref{eq:flowequations}):
\begin{eqnarray}
\eta &=& \frac{1}{2}\left(2 \epsilon + s\right),\cr
{}^2 \lambda &=&= \frac{1}{2}\left(2 \epsilon + s\right) \left(\epsilon + s\right),\cr
{}^{\ell + 1}\lambda &=& {}^\ell \lambda \left[\epsilon + \frac{1}{2}\left(\ell + 1\right) s\right],\cr
\rho &=& \frac{3 s^2}{2 \epsilon},\cr
{}^2 \alpha &=& \frac{3 s^3}{\epsilon},\cr
{}^{\ell + 1} \alpha &=& \frac{1}{2}\left(\ell + 3\right) s \left({}^\ell \alpha\right).
\end{eqnarray}
In this case, the sound horizon $c_S H^{-1}$ and the Hubble length $H^{-1}$ evolve independently. We can solve for their behavior as functions of the field $\phi$ as follows:
First, from $\epsilon = {\rm const.}$, 
\begin{equation}
s \equiv \frac{2 M_P^2}{\gamma} \left(\frac{H'}{H}\right) \frac{\gamma'}{\gamma} = \pm M_p \sqrt{2 \epsilon} \frac{\gamma'}{\gamma^{3/2}} = {\rm const.}
\end{equation}
so that 
\begin{equation}
\frac{d \gamma}{\gamma^{3/2}} = \pm \frac{s}{M_p \sqrt{2 \epsilon}} d\phi,
\end{equation}
with solution
\begin{equation}
\label{eq:gammaNS}
\gamma = \left(\frac{8 M_p^2 \epsilon}{s^2}\right)\frac{1}{\phi^2},
\end{equation}
where we have chosen $\gamma \rightarrow \infty$ (or $c_S = 0$), corresponding to the ``tip'' of the warped throat, to be $\phi = 0$. 
We can solve for the dependence of the field $\phi$ on the number of e-folds $N$ from $d s / d N = 0$, which implies $\gamma \propto e^{s N}$, and from Eq. (\ref{eq:gammaNS}) we have
\begin{equation}
\phi^2 \propto e^{- s N}.
\end{equation}
Note in particular that while $\epsilon$ is manifestly positive, $s$ can have either sign, and the direction of the field evolution depends on the sign of $s$:
\begin{equation}
\frac{d \phi}{\phi} = - \frac{s}{2} d N.
\end{equation}
Having obtained $\gamma\left(\phi\right)$, we can solve for $H\left(\phi\right)$ by using the definition of $\epsilon$,
\begin{equation}
\epsilon \equiv \frac{2 M_P^2}{\gamma} \left(\frac{H'}{H}\right)^2 = \frac{s^2 \phi^2}{4 \epsilon}  \left(\frac{H'}{H}\right)^2 = {\rm const.},
\end{equation}
and find that
\begin{equation}
\frac{d H}{H} = \pm \frac{2 \epsilon}{s} \frac{d \phi}{\phi},
\end{equation}
with solution
\begin{equation}
H \propto \phi^{\pm 2 \epsilon / s}.
\end{equation}
The sign ambiguity can be resolved by requiring that the universe be inflating, {\it i.e.} $d H / d N > 0$, so that
\begin{equation}
\label{eq:Hphi}
H \propto \phi^{- 2 \epsilon / s} \propto e^{\epsilon N}.
\end{equation}
Since $d N < 0$ for $dt > 0$, we see that $s < 0$ corresponds to a field evolving toward the tip of the throat at $\phi = 0$, referred to in the literature as a {\it UV} model \cite{Chen:2006nt,Lidsey:2007gq}. Conversely, $s > 0$ corresponds to a field evolving away from the tip of the throat, or an {\it IR} model \cite{Chen:2004gc,Chen:2005ad}. For example, the AdS model considered by Alishahiha, {\it et al.} \cite{Alishahiha:2004eh} corresponds to the case $s = - 2 \epsilon$. For $\epsilon$ constant, inflation formally continues forever into both the past and the future, so we must introduce additional physics from a particular realization of DBI to determine the field value which corresponds to the end of inflation at $N = 0$. In the case of a UV model, inflation ends with brane collision near the tip of the throat at some $\phi_0 > 0$. In the case of an IR model, we see from Eq. (\ref{eq:cS}) that the speed of sound is increasing as the field evolves, and will eventually exceed unity for a finite number
of e-folds for any choice of initial condition. To interpret this behavior, it is useful to rewrite Eqs. (\ref{eq:Hamjacobi}) for $V\left(\phi\right)$ and $f\left(\phi\right)$ corresponding to the case of constant $\epsilon$ and $s$ as follows: 
\begin{eqnarray}
\label{eq:potentials}
V\left(\phi\right) &=& 3 M_P^2 H^2\left(\phi\right) \left[1 - \left(\frac{2 \epsilon}{3}\right) \frac{1}{1 + c_S\left(\phi\right)}\right],\cr
f\left(\phi\right) &=& \left(\frac{1}{2 M_P^2 \epsilon}\right) \frac{1 - c_S^2\left(\phi\right)}{H^2\left(\phi\right) c_S\left(\phi\right)},
\end{eqnarray}
where $H\left(\phi\right)$ is given by Eq. (\ref{eq:Hphi}), and 
\begin{equation}
c_S \propto \phi^2 \propto e^{-s N}.
\end{equation}
While the potential $V\left(\phi\right)$ is well-behaved when $c_S > 1$, the function $f\left(\phi\right)$ becomes negative. For a metric of the form (\ref{eq:DBImetric}), with $h\left(\phi\right)$ given by Eq. (\ref{eq:warpfactor}), we see immediately that $f(\phi) < 0$ corresponds to an imaginary $h\left(\phi\right)$, indicating a breakdown of 
the relation (\ref{eq:warpfactor}) at the point where the throat joins onto the bulk manifold. For simplicity, we will assume that our solution is valid close to the end of the throat, which we will take to be at the point where $c_S = 1$ and the 6-dimensional bulk metric (\ref{eq:DBImetric}) will have a more general form which will not necessarily be well-parameterized by a single radial coordinate $r$. We will also take this to define the end of inflation for $s > 0$ (IR) case, and the beginning of inflation in the $s < 0$ (UV) case. We then have a well-defined solution for the evolution in the throat, with $V\left(\phi\right)$ and $f\left(\phi\right)$ given by Eqs. (\ref{eq:potentials}), and
\begin{equation}
\phi^2 = \phi_0^2 e^{-s N}.
\end{equation}
We can then write the expansion rate and sound speed as
\begin{eqnarray}
\label{eq:nonscaling}
H\left(\phi\right) &=& H_0 \left(\frac{\phi}{\phi_0}\right)^{- 2 \epsilon / s} = H_0 e^{\epsilon N},\cr
c_S\left(\phi\right) &=& c_S\left(\phi_0\right) \left(\frac{\phi}{\phi_0}\right)^2 = c_S\left(\phi_0\right) e^{-s N},
\end{eqnarray}
where $c_S\left(\phi_0\right) \ll 1$ in the $s < 0$ case, and $c_S\left(\phi_0\right) \simeq 1$ in the $s > 0$ case, so that Eq. (\ref{eq:gammaNS}) becomes
\begin{equation}
\phi_0^2 \simeq \frac{8 M_P^2 \epsilon}{s^2}.
\end{equation}
We can derive the time dependence of the field $\phi$ and scale factor $a$ from
\begin{equation}
\dot\phi = \pm \sqrt{\frac{2 \epsilon}{\gamma}} M_P H\left(\phi\right),
\end{equation}
so that, using Eq. (\ref{eq:gammaNS}) and Eq. (\ref{eq:nonscaling}), we have
\begin{equation}
t = \frac{1}{H_0 \epsilon} \left(\frac{\phi}{\phi_0}\right)^{2 \epsilon / s} = \frac{1}{\epsilon H\left(t\right)}.
\end{equation}
The field therefore evolves as
\begin{equation}
\phi\left(t\right) = \phi_0 \left(\epsilon H_0 t\right)^{s / 2 \epsilon},
\end{equation}
and the scale factor is again a power-law in time,
\begin{equation}
a\left(t\right) \propto t^{1 / \epsilon}.
\end{equation}
This corresponds exactly to the solution of Chimento and Lazkoz \cite{Chimento:2007es}. 

In the next section, we derive an exact equation for quantum fluctuations which can be solved analytically for the above background solutions.

\section{Solutions for quantum modes}
\label{sec:modes}

Inflationary spacetimes have the property that quantum fluctuations on very small scales grow in wavelength due to expansion at a rate faster than the growth of the horizon size, so that they dynamically evolve to superhorizon scales, where they ``freeze'' as classical perturbations \cite{Mukhanov:1981xt,Hawking:1982my,Starobinsky:1982ee,Guth:1982ec,Bardeen:1983qw}. Quantum perturbations generated during inflation are of two types: scalar (or curvature) perturbations, and tensor (or gravitational wave) perturbations. We are primarily interested in curvature perturbations, because these perturbations are sensitive to the sound horizon, whereas tensor perturbations are not. Furthermore, production of tensors in DBI inflation is generically suppressed \cite{Baumann:2006cd,Lidsey:2007gq,Kobayashi:2007hm}, although versions involving multiple branes can produce significant tensors \cite{Krause:2007jr,Becker:2007ui,Huang:2007hh,Ward:2007gs}. Generation of perturbations 
in scenarios with a general speed of sound were treated in detail by Garriga and Mukhanov \cite{Garriga:1999vw}, who showed that the power spectrum of curvature perturbations is given by
\begin{equation}
\label{eq: scalar spectrum}
P_{\cal R}\left(k\right) = \frac{k^3}{2 \pi^2} \left\vert\frac{u_k}{z}\right\vert^2,
\end{equation}
where $k$ is a comoving wave number and the quantum mode function $u_k$ satisfies
\begin{equation}
\label{eq:uk}
u_k'' + \left(c_s^2 k^2 - \frac{z''}{z}\right) u_k = 0.
\end{equation}
Here a prime denotes a derivative with respect to conformal time $d\tau \equiv dt / a$, and $z$ is given by
\begin{equation}
z \equiv \frac{a \sqrt{\rho + p}}{c_S H} = \frac{a \gamma^{3/2} \dot\phi}{H} = - M_p a \gamma \sqrt{2 \epsilon}.
\end{equation}
In order to solve Eq. (\ref{eq:uk}) for the solutions derived in Sec. \ref{sec:background}, we need to obtain an expression for $z''/z$ in terms of the parameters $\epsilon$ and $s$. We first derive a general expression, and then specialize to the case where $\epsilon$ and $s$ are constant. Writing 
\begin{equation}
\frac{d}{d \tau} = a \frac{d}{d t} = - a H \frac{d}{d N},
\end{equation}
we have, using the flow equations (\ref{eq:flowequations}),
\begin{eqnarray}
z' &=& M_P \sqrt{2} a H \frac{d}{d N} \left(a \gamma \sqrt{\epsilon}\right)\cr
&=& - M_P \sqrt{2} a^2 H \gamma \sqrt{\epsilon} \left(1 + \epsilon - \eta - \frac{1}{2}s \right).
\end{eqnarray}
Similarly, it is straightforward to show that
\begin{equation}
\frac{z''}{z} = a^2 H^2 F\left(\epsilon,\eta,s,\rho,{}^2\lambda\right),
\end{equation}
where \cite{Shandera:2006ax}
\begin{eqnarray}
\label{eq:F}
F = &&2 + 2 \epsilon - 3 \eta - \frac{3}{2} s - 4 \epsilon \eta + \frac{1}{2} \eta s  + 2 \epsilon^2\cr
 &&+ \eta^2 - \frac{3}{4} s^2 + {}^2\lambda + \frac{1}{2}\epsilon\rho.
\end{eqnarray}
This expression is exact, and involves no assumption of slow roll. 

It is useful to change the time variable in the mode equation (\ref{eq:uk}) to the ratio of the wavenumber to the sound horizon,
\begin{equation}
y \equiv \frac{k}{\gamma a H},
\end{equation}
so that
\begin{equation}
\frac{d}{d\tau} = - a H \frac{d y}{dN}\frac{d}{d y} = - a H \left(1 - \epsilon - s\right) y \frac{d}{dy},
\end{equation}
and
\begin{equation}
\frac{d^2}{d\tau^2} = a^2 H^2 \left[\left(1 - \epsilon - s\right)^2 y^2 \frac{d^2}{dy^2} + G\left(\epsilon,\eta,s,\rho\right) y \frac{d}{d y}\right],
\end{equation}
where
\begin{eqnarray}
\label{eq:G}
G &=& \gamma \frac{d}{dN} \left[\frac{1}{\gamma} \left(1 - \epsilon - s\right)\right]\cr
&=& -s + 3 \epsilon s - 2 \epsilon \eta - \eta s + 2\epsilon^2 + 3 s^2 - \epsilon \rho.
\end{eqnarray}
The exact mode equation (\ref{eq:uk}) is then
\begin{equation}
\label{eq:exactmode}
\left(1 - \epsilon -s\right)^2 y^2 \frac{d^2 u_k}{dy^2} + G y \frac{d u_k}{d y} + \left[y^2 - F\right] u_k = 0,
\end{equation}
where $F$ and $G$ are purely functions of flow parameters, given by Eqs. (\ref{eq:F}) and (\ref{eq:G}), respectively.

We next solve the mode equation for the case of constant DBI flow parameters. Following the standard definition for the adiabatic vacuum, we have that the solutions to the mode equation (\ref{eq:uk}) are given by the WKB solution
\begin{equation}
\label{eq:general solution}
u_{k}=\frac{N_{k}}{\sqrt{c_{s}}}\exp{\left(\pm ik \int^{\tau}c_{s}d\tau \right)},
\end{equation}
assuming that the quantum modes evolve much faster than the background expansion. The adiabatic vacuum is an excellent approximation of the Minkowski vacuum in the limit where the background expansion is negligible relative to the frequency of the quantum fluctuations. The above condition can quantitatively be expressed by 
\begin{equation}
\label{eq:danielsson's condition}
\frac{d}{d \tau}\left(\frac{z''}{z}\right) \ll \omega,
\end{equation}
where
\begin{equation}
\label{eq:omega}
\omega=\sqrt{(c_{s}k)^2-z''/z}.
\end{equation}
In terms of $y$, condition (\ref{eq:danielsson's condition}) becomes
\begin{equation}
\label{eq:condition in terms of y 1}
(c_{s}k)^2 \gg (aH)^2\left(6+\mathcal{O}(\epsilon,\eta,s)+...\right),
\end{equation}
or
\begin{equation}
\label{eq:condition in terms of y 2}
y \gg 1.
\end{equation}
The above condition can therefore be expressed equivalently as the small wavelength limit of Eq. (\ref{eq:uk}) and in this limit the quantum modes do not feel the expansion of the universe. The modes $u_{k}$ can be normalized using the canonical quantization condition for the fluctuations
\begin{equation}
\label{eq:Wronskian}
u_{k}^{*}\frac{du_{k}}{d \tau}-u_{k}\frac{du_{k}^{*}}{d \tau}=-i,
\end{equation} 
and then
\begin{equation}
\label{eq:general solution 2}
u_{k}=\frac{1}{\sqrt{2c_{s}k}}\exp{\left(\pm ik \int^{\tau}c_{s}d\tau \right)}.
\end{equation}
It should be noted that the above expression is a general result, independent of the background evolution. 

If we now impose the specific background evolution which corresponds to constant DBI flow parameters, and use the following exact expression
\begin{equation}
\label{eq:dy dtau}
dy=-c_{s}k(1-\epsilon-s)d\tau,
\end{equation}
we find that in the small wavelength limit
\begin{equation}
\label{eq:modes for large y}
u_{k}=\frac{1}{\sqrt{2c_{s}k}}e^{iy/(1-\epsilon-s)},
\end{equation}
keeping the positive frequency modes only. The evolution of $u_{k}$ at all times can be found by solving the mode equation (\ref{eq:exactmode}) expressed in terms of $y$. For the case considered here, the mode equation reduces to
\begin{eqnarray}
\label{eq:mode equation for constant parms. 1}
(&1-\epsilon-s&)^2y^2\frac{d^2u_{k}}{dy^2}+s(-1+\epsilon+s)y\frac{du_{k}}{dy}\cr&&+[y^2-(1-s)(2-s-\epsilon)]u_{k}=0,
\end{eqnarray}
with solutions proportional to Hankel functions of the first and second kind
\begin{eqnarray}
\label{eq:solution of mode eqaution for constant parms. 1}
u_{k}(y)&=&y^{\frac{1-\epsilon}{2(1-\epsilon-s)}}\left[C_{1}H_{\nu}^{(1)}\left(\frac{y}{1-\epsilon-s}\right)\right.\cr&&\left.+C_{2}H_{\nu}^{(2)}\left(\frac{y}{1-\epsilon-s}\right)\right],
\end{eqnarray}
where
\begin{equation}
\label{eq:nu for mode equation for constant parms. 1}
\nu= \frac{3-2s-\epsilon}{2(1-\epsilon-s)}.
\end{equation}
Imposing the Bunch-Davies vacuum by setting $C_2=0$ and using that 
\begin{equation}
\label{eq:cs y}
c_{s}=By^b,
\end{equation}
where $B$ and $b$ are constants (see Sec. \ref{sec:vacuum}) we can rewrite Eq. (\ref{eq:solution of mode eqaution for constant parms. 1}) as 
\begin{equation}
\label{eq:solution of mode eqaution for constant parms. 2}
u_{k}(y)=C_{1}\sqrt{\frac{B}{c_{s}}}\sqrt{y}H_{\nu}\left(\frac{y}{1-\epsilon-s}\right).
\end{equation}
By expressing the Wronskian condition (\ref{eq:Wronskian}) in terms of $y$ as
\begin{equation}
\label{eq:Wronskian y}
u_{k}^{*}\frac{du_{k}}{dy}-u_{k}\frac{du_{k}^{*}}{dy}=\frac{i}{c_{s}k(1-\epsilon-s)},
\end{equation} 
it can be shown that the normalization of the modes $C_{1}\sqrt{B}$ is given by
\begin{equation}
\label{eq:normalization}
C_{1}\sqrt{B}=\sqrt{\frac{\pi}{4k(1-\epsilon-s)}}.
\end{equation}
Equation (\ref{eq:solution of mode eqaution for constant parms. 2}) then finally becomes
\begin{equation}
\label{eq:solution of mode eqaution for constant parms. 3}
u_{k}(y)=\frac{1}{2}\sqrt{\frac{\pi}{c_{s}k}}\sqrt{\frac{y}{1-\epsilon-s}}H_{\nu}\left(\frac{y}{1-\epsilon-s}\right),
\end{equation}
where it can easily be seen that it reduces to Eq. (\ref{eq:modes for large y}) for large $y$, thus giving the expected behavior in the small wavelength limit.

We next calculate the power spectrum of scalar perturbations $P_{\mathcal{R}}^{1/2}(k)$. Taking the long wavelength limit of Eq. (\ref{eq:solution of mode eqaution for constant parms. 3}) we find that
\begin{equation}
\label{eq:solution of mode eqaution for constant parms. at small y}
\left|u_{k}(y)\right| \rightarrow 2^{\nu-3/2}\frac{\Gamma(\nu)}{\Gamma(3/2)}(1-\epsilon-s)^{\nu-1/2} \frac{y^{1/2-\nu}}{\sqrt{2 c_{s}k}},
\end{equation}
and then
\begin{equation}
\label{eq:scalar spectrum 2}
P_{\mathcal{R}}^{1/2}=\mathcal{V}(\nu)\left.\frac{H^2}{2\pi \dot\phi}\right|_{k=\gamma a H},
\end{equation}
where the fluctuation amplitude is evaluated at horizon crossing and 
\begin{equation}
\label{eq:mathcalV}
\mathcal{V}(\nu)=2^{\nu-3/2}\frac{\Gamma(\nu)}{\Gamma(3/2)}(1-\epsilon-s)^{\nu-1/2}= {\rm const.}
\end{equation}
Since the DBI flow parameters are constant, the expression for the spectral index, $n_{s}$ defined by
\begin{equation}
\label{eq:spec index 1}
n_{s}-1 \equiv \frac{d(lnP_{\mathcal{R}})}{d(lnk)},
\end{equation}
simplifies considerably. Using that 
\begin{equation}
\label{eq:dlnk vs dN}
\frac{d}{d(lnk)}=-\left(\frac{1}{1-\epsilon-s}\right)\frac{d}{dN}
\end{equation}
at horizon crossing, Eq. (\ref{eq:spec index 1}) becomes
\begin{equation}
\label{eq:spec index 2}
n_{s}=1-\frac{2\epsilon+s}{1-\epsilon-s}.
\end{equation}
The above expression for $n_{s}$ is exact and it reduces to the well known result for power law inflation of a canonical field in the $s=0$ limit. 

We can generalize the preceding analysis to the case where the flow parameters are small and can vary slowly. The mode equation (\ref{eq:exactmode}) to first order in the flow parameters reduces then to
\begin{eqnarray}
\label{eq:exactmode 2}
(1&- 2\epsilon &-2s)y^2 \frac{d^2 u_k}{dy^2}-s y \frac{d u_k}{d y}\cr&& + \left[y^2 - 2(1+\epsilon-\frac{3}{2}\eta-\frac{3}{4}s)\right] u_k = 0.
\end{eqnarray}
The solutions will again be given by Eq. (\ref{eq:solution of mode eqaution for constant parms. 3}), but the order of the Hankel function will now have an explicit dependence on $\eta$  as follows
\begin{equation}
\label{eq:nu 2}
\nu=\frac{3}{2}+2\epsilon-\eta+s.
\end{equation}
Combining Eqs. (\ref{eq:scalar spectrum 2}) and (\ref{eq:mathcalV}) we find that the power spectrum of scalar perturbations $P_{\mathcal{R}}^{1/2}$ to lowest order in $\epsilon$, $\eta$ and $s$ is given by
\begin{eqnarray}
\label{eq:scalar spectrum 3}
P_{\mathcal{R}}^{1/2}&=&[(1-\epsilon-s)\cr&&+(2-ln2-\gamma)(2\epsilon-\eta+s)]\left.\frac{H^2}{2\pi \dot\phi}\right|_{k=\gamma a H}
\end{eqnarray}
where $\gamma \approx 0.577$ is Euler's constant. Using finally that 
\begin{equation}
\label{eq:spec index horizon crossing}
n_{s}-1=\frac{-2}{1-\epsilon-s}\frac{d(lnP_{\mathcal{R}}^{1/2})}{dN}
\end{equation}
at horizon crossing, we find an expression for the spectral index to second order in the flow parameters,
\begin{eqnarray}
\label{eq:spec index second order}
n_{s}&=&1-4\epsilon+2\eta-2s-2(1+C)\epsilon^2-(3+C)s^2\cr &&-\frac{1}{2}(3-5C)\epsilon\eta-\frac{1}{2}(11+3C)\epsilon s+(1+C)\eta s\cr&& +\frac{1}{2}(1+C)\epsilon \rho +\frac{1}{2}(3-C) ({}^2 \lambda),
\end{eqnarray}
where
\begin{equation}
\label{eq:C}
C=4(ln2+\gamma)-5.
\end{equation}
Equation (\ref{eq:spec index second order}) is a generalization of Stewart and Lyth's result \cite{Stewart:1993bc} for the case of inflation with a slowly time-varying speed of sound. Note that this expression differs at second order from the similar expression derived by Shandera and Tye \cite{Shandera:2006ax,Bean:2007hc}, who did not perform a fully self-consistent expansion of the mode equation in the flow parameters. Our result agrees with those derived in Refs. \cite{Wei:2004xx,Chen:2006nt}. We emphasize that the results derived in this section apply to {\it generic} k-inflation models, not just DBI. 

\section{Issues for vacuum selection}
\label{sec:vacuum}

The solution for the power spectrum derived in Sec. \ref{sec:modes} depends strongly on the assumption that it is possible to self-consistently take the initial quantum state of the mode to be the Bunch-Davies vacuum, which for a varying speed of sound corresponds to a WKB state,
\begin{equation}
u_k \propto \frac{1}{\sqrt{c_S}} \exp{\left(-i k \int{ c_s d\tau}\right)},
\end{equation}
defined in the limit that $-k \tau \propto y \rightarrow \infty$. In this section, we examine the validity of this assumption. Using $d N = - a H d\tau$, we can write the parameter $s$ as
\begin{equation}
s \equiv \frac{1}{\gamma}\frac{d \gamma}{d N} = \frac{1}{a H} \left(\frac{1}{c_s}\frac{d c_s}{d \tau}\right) = \frac{1}{a H} \frac{d y}{d \tau} \left(\frac{1}{c_s}\frac{d c_s}{d y}\right).
\end{equation}
Using
\begin{equation}
\frac{d y}{d \tau} = c_S k \left(\epsilon + s - 1\right),
\end{equation}
we have the relation
\begin{equation}
\frac{y}{c_S}\frac{d c_s}{d y} = \frac{s}{\epsilon + s - 1} \equiv b = {\rm const.},
\end{equation}
so that we can express the speed of sound as a function of $y$ for a given mode:
\begin{equation}
\label{eq:koumo}
c_S \propto y^b.
\end{equation}
From the solution (\ref{eq:nonscaling}), $c_S \propto e^{-s N}$, and therefore
\begin{equation}
\label{eq:yevol}
y \propto e^{N \left(1 - \epsilon - s\right)}.
\end{equation}
For a Bunch-Davies vacuum to exist, we must have $y \rightarrow \infty$ at early times, {\it i.e.} $N \rightarrow \infty$, which results in a condition on the flow parameters
\begin{equation}
\label{eq:BDcondition}
\epsilon < 1 - s.
\end{equation}
For the $s > 0$ (IR) case, this is a stronger condition than that required for inflation to occur, which is $\epsilon < 1$. Therefore, IR models admit solutions for which inflation occurs, but quantum modes never reach the Bunch-Davies limit. The physical basis for this is clear from considering the behavior of the sound horizon
\begin{equation}
c_s H^{-1} \propto e^{- \left(\epsilon + s\right) N},
\end{equation}
compared to the scale factor $a \propto e^{-N}$. For $\epsilon > 1 - s$, the sound horizon expands more quickly than the scale factor, and modes fall {\it into} the sound horizon, that is, they evolve from superhorizon scales at early time to subhorizon scales at late time. Since $\epsilon$ is identically positive, the Bunch-Davies vacuum does not exist for {\it any} model with $s \geq 1$. 

The UV case, with $s < 0$, is more subtle. As long as the space is inflating, the condition (\ref{eq:BDcondition}) is always satisfied, and $y \rightarrow \infty$ as $N \rightarrow \infty$. However, $c_S \rightarrow \infty$ in this limit as well, as it can be seen from Eq. (\ref{eq:koumo}), and we do not have a well-defined 4-D effective theory within which to set a vacuum state. Since inflation only lasts for a finite period, we never reach the asymptotically short-wavelength limit. However, we can still define a nearly Bunch-Davies WKB state as long as all modes of astrophysical interest have wavelengths much smaller than the sound horizon at the onset of inflation, which we take to be at $c_S = 1$. We take inflation to end via brane collisions near the tip of the throat at $\phi = \phi_0 > 0$, for which $c_S\left(\phi_0\right) \ll 1$. The total number of e-folds of inflation $N_{\rm tot}$ is then given by the condition
\begin{equation}
c_S\left(\phi_0\right) e^{-s N_{\rm tot}} = 1,
\end{equation}
or 
\begin{equation}
\label{eq:limit}
c_S\left(N\right) = e^{s \left(N_{\rm tot} - N\right)}.
\end{equation}

The sound speed is related to the primordial non-Gaussianity by \cite{Chen:2006nt}
\begin{equation}
f_{NL}^{\rm equil} = \frac{35}{108} \left(\frac{1}{c_S^2} -1 \right) < 332,
\end{equation}
where the upper limit on the parameter $f_{NL}^{\rm equil}$ for equilateral modes is from the WMAP data \cite{Creminelli:2006rz}. This places a lower limit on the sound speed of $c_S > 0.03$ at CMB scales, and using Eq. (\ref{eq:limit}) this condition can be written as
\begin{equation}
\label{eq:limit2}
s \left(N_{\rm tot} - N_{\rm hor}\right)>-3.5,
\end{equation}
where we define $N_{\rm hor}$ to be the number of e-folds when scales of order the horizon size today exited the horizon during inflation. In the absence of late-time entropy production in the universe, this number is in the range $N_{\rm hor} = [46,60]$, assuming the reheat temperature must be at least at the electroweak scale $T_{\rm RH} > 1\ {\rm TeV}$  \cite{Liddle:2003as,Kinney:2005in}. This results in an upper limit on the total number of e-folds,
\begin{equation}
\label{eq:NGlimit}
N_{\rm tot} - N_{\rm hor} < \left\vert\frac{3.5}{s}\right\vert.
\end{equation}
Note that this is a {\it very} rough limit, since the speed of sound, and therefore the level of non-Gaussianity, is scale-dependent.  This is not taken into account in the WMAP limit on $f_{NL}$. (See Ref. \cite{LoVerde:2007ri} for a discussion on scale-dependent non-Gaussianity.) Similarly, the existence of a Bunch-Davies vacuum puts a lower limit on the total number of e-folds, as follows: Using Eq. (\ref{eq:yevol}), we can write the value $y_i$ at the onset of inflation for a mode $k$ as 
\begin{equation}
y_i\left(k\right) = e^{\left[N_{\rm tot} - N_{\rm hor}\left(k\right)\right] \left(1 - \epsilon - s\right)} 
\end{equation}
where $N_{\rm hor}\left(k\right)$ is defined to be the number of e-folds when the mode with wavenumber $k$ crossed the horizon. We can consider a Bunch-Davies vacuum to exist for all modes of astrophysical interest if for all modes $k$ inside our current horizon, $y_i$ is sufficiently large that finite-wavelength effects are unobservable, $y_i \geq \mathcal{O}(10^3)$ \cite{Easther:2004vq}. We therefore have a condition for a valid Bunch-Davies state
\begin{equation}
\label{eq:Ntotbound}
N_{\rm tot} - N_{\rm hor} > \frac{6.9}{\left(1 - \epsilon - s\right)}.
\end{equation}
For models such as that of Alishahiha, {\it et al.} \cite{Alishahiha:2004eh} with a nearly scale-invariant spectrum, $s \simeq -2 \epsilon$, the bounds (\ref{eq:NGlimit}) and (\ref{eq:Ntotbound}) are comparable for $s \sim -0.7$.  For $s < -0.7$, the total number of e-folds is too small to guarantee a Bunch-Davies initial state for all modes. Therefore the primordial power spectrum on large scales will necessarily be sensitive to physics outside of the throat, since the initial vacuum state must be set in bulk space. In this circumstance, the spectrum of primordial density perturbations will in general be sensitive to the physics of the bulk space, possibly providing an observationally accessible signal of stringy physics. Note especially that a more stringent upper bound on the amplitude of non-Gaussianity from upcoming measurements would strengthen the upper bound (\ref{eq:NGlimit}), and therefore broaden the parameter range for 
which observation is in principle sensitive to bulk physics.\footnote{The recent claimed detection of non-Gaussianity by Yadav and Wandelt \cite{Yadav:2007yy} is for the case of squeezed triangles, not equilateral triangles. If confirmed, such a detection will rule out  single-field slow-roll inflation models as well as DBI \cite{Chen:2006nt}.}  A determination of the specific form (if any) of an observational signature will require analysis within the context of a well-defined stringy model, where it is possible to calculate the vacuum state in the bulk space.

\section{Conclusions}
\label{sec:Conclusions}

In this paper we have considered a family of k-inflation models with exact solutions for the background cosmological evolution characterized by power-law expansion, $a(t) \propto t^{1/\epsilon}$, where $\epsilon = {\rm const.}$ is the first slow roll parameter. The set of solutions is parameterized by the flow parameters $\epsilon$ and $s$, such that the evolution of the Hubble parameter in terms of the number of e-folds $N$ is
\begin{equation}
H \propto e^{\epsilon N},
\end{equation}
and the evolution of the sound speed is
\begin{equation}
c_S \propto e^{- s N}.
\end{equation}
In the case of DBI inflation, these solutions correspond to particular choices of the potential $V\left(\phi\right)$ and the inverse brane tension $f\left(\phi\right)$, given by Eqs. (\ref{eq:Vscaling}) and (\ref{eq:fscaling}) for $s = 0$, and by Eqs. (\ref{eq:potentials}) for $s \neq 0$. The equation for the evolution of quantum modes can be solved exactly in the case of power-law inflation, with the spectral index for scalar fluctuations given by
\begin{equation}
n_{s}=1-\frac{2\epsilon+s}{1-\epsilon-s}.
\end{equation}
We also consider the general slow-roll case, for which the inflationary flow parameters are small and approximately constant. In this case the scalar spectral index is given to second order in the flow parameters by:
\begin{eqnarray}
n_{s}&=&1-4\epsilon+2\eta-2s-2(1+C)\epsilon^2-(3+C)s^2\cr &&-\frac{1}{2}(3-5C)\epsilon\eta-\frac{1}{2}(11+3C)\epsilon s+(1+C)\eta s\cr&& +\frac{1}{2}(1+C)\epsilon \rho +\frac{1}{2}(3-C) ({}^2 \lambda),
\end{eqnarray}
where $C = 4 \left(\ln{2} + \gamma\right) - 5 \simeq 0.08$. This differs slightly from the expression derived in Ref. \cite{Shandera:2006ax}, and is a more accurate second-order expression. 

Of particular importance for the self-consistency of this solution is the behavior of comoving scales relative to the sound horizon as expressed by $y\equiv c_S k / \left(a H\right)$, which in terms of the number of e-folds $N$ is given by
\begin{equation}
y \propto e^{\left(1 - \epsilon - s\right) N}.
\end{equation}
For the short-wavelength limit $y \rightarrow \infty$ to exist, the background evolution must satisfy the condition $\epsilon < 1 - s$, which is always satisfied for $s < 0$, but is a nontrivial condition for $s > 0$. If $\epsilon > 1 - s$, comoving modes are superhorizon at early times, and subhorizon at late times, and therefore it is not possible to define an initial condition for the modes in the short-wavelength limit. In the case of DBI inflation, the $s > 0$ case corresponds to the brane rolling out of the warped throat (IR models), and $s < 0$ corresponds to the brane rolling into the warped throat (UV models). In the $s < 0$ case, observational constraints on primordial non-Gaussianity imply an upper limit on the total number of e-folds of inflation of $N_{\rm tot} - N_{\rm hor}< \left\vert 3.5 / s\right\vert$, but the existence of a well-defined Bunch-Davies vacuum places a lower limit on the total number of e-folds of $N_{\rm tot} - N_{\rm hor} > 6.9 / \left(1 - \epsilon - s\right)$. For models in which the total number of e-folds is close to (or less than) this limit, the initial state of the quantum modes in inflation will in general be sensitive to physics in the bulk space, not just the throat. More stringent upper bounds on primordial non-Gaussianity from upcoming observations would make this tension more severe. This opens up the possibility that the spectrum of primordial density perturbations in DBI inflation can probe the physics of the higher-dimensional bulk space as well as that of the warped throat.

\begin{center}
{\bf ACKNOWLEDGMENTS}
\end{center}
We thank Niayesh Afshordi, Ghazal Geshnizjani, Eric Greenwood, Justin Khoury, Robert Myers, Claudia de Rahm, Andrew Tolley, John Wang, and Mark Wyman for helpful discussions. We thank Sarah Shandera for comments on a draft version of this paper. This  research is supported  in part  by  the National  Science  Foundation under  grant NSF-PHY-0456777.

\end{document}